# Digital twins' kinetics of virtual free-radical copolymerization of vinyl monomers with stable radicals. 2. Styrene

Elena F. Sheka

Institute of Physical Researches and Technology, Peoples' Friendship University of Russia (RUDN University), 117198 Moscow, Russia;

sheka@icp.ac.ru

**Abstract:** The first experience of virtual free-radical copolymerization (FRCP) of methyl methacrylate with stable radicals fullerene $C_{60}$ and *TEMPO* in the framework of the digital twins (DTs) concept (arXiv:2309.11616 [physics.chem-ph]) is extended to styrene. The virtualization of the chemical process is based on a conceptual view of this process from the perspective of a chain reaction that covers a set of elementary reactions (ERs). Such an approach is the most suitable for quantum chemical treatment. The calculations, concerning about 50 such ERs as well as associated DTs, were performed using a semi-empirical version of the unrestricted Hartree-Fock approximation. The main energy and spin-density parameters of the DTs' ground state are determined. The barrier profiles of the selected DTs were calculated for the activation energy of the studied reactions to be determined. The decisive role of spins in the formation of transition states of these processes was confirmed. The two stable radicals behave quite differently with respect to the standard FRP of styrene. *TEMPO* effectively captures monomer-radicals of styrene with the acting free radical not depending on which namely radical is in use. Its copolymerizing action manifests itself as a 'killing' of the FRP thus providing the appearance of an induction period (IP) in the total process kinetics. In contrast, the copolymerizing action of fullerene drastically depends on the active free radical. In the case of the alkyl-nitrile one, the fullerene switches to the role of initiator and first provides the styrene polymerization followed with anchoring of the formed polymer chains to the molecule body, after which the standard FRP starts. In the case of benzoyl-peroxide one, fullerene acts an inhibitor thus capturing formed monomer-radicals. The obtained virtual kinetic data are in a full consent with experimental reality.

**1. Introduction**

A general idea of polymerization as a chain reaction [1,2], approaching its century, is experiencing a new birth. Chain reaction, as a set of independent elementary reactions (ERs), has proven to be in high demand in modern virtual polymerics based on the concept of digital twins (DTs) [3,4]. The latter implements the long-desired consideration of reactions of the same type, carried out under the same conditions, which makes it possible to determine with high conviction the main trends accompanying a complex and multifaceted polymerization process [5-7]. Recently published first results [3,4] are related to the virtualization of free-radical polymerization (FRP) of vinyl monomers and the role in this process of small additives of stable radicals such as TEMPO and $C_{60}$ fullerene. They made it possible to see the outlines of not only the full contour of these processes, but also to notice their specific features, which previously remained in shadows. In this work, we continue going in this direction and move on to the consideration of the most studied FRP of organic monomers—FRP of styrene [8-23].

A general algorithm of the DT concept [24,25] is described by a well known trinity of digitalization actions that are

*Digital twins → Virtual device → IT product*.

Here DTs are molecular models under study; virtual device presents a carrier of the selected computational soft, while IT product covers a large set of structural, thermodynamic and kinetic virtual results related to the studied case. Digital twins of the current study are associated with ERs that govern the FRP initial stage. Their composition is established with the relevant ERs' set, depending on concrete participants of the chemical process [4, 17]. In the current study, the choice of the related ERs set was stimulated with fundamental empirical studies [22, 23] concerning the FRP of styrene, initiated with either alkyl-nitrile $AIBN^\bullet$ or benzoyl-peroxide $BP^\bullet$ free radicals produced in the course of the thermal decomposition of 2,2'-azobisisobutyronitrile (AIBN) and benzoyl peroxide (BP), respectively. Additionally subjected to the action of small additives of two stable radicals fullerene $C_{60}$ and TEMPO, FRP is expended to FRCP. A rather complete list of the relevant ERs, simultaneously nominating DTs of their final products, is presented in Table 1. As seen in the table, the number of the corresponding objects is quite large so that the digitalization of the considered chemical process seems to be convincing enough.

Virtual device in the current study is the CLUSTER-Z1 software [26, 27] implementing AM1 version of the semi-empirical unrestricted two-determinant Hartree-Fock (UHF) approach [28]. The program showed itself highly efficient concerning open-shell electronic systems such as fullerenes [29,30], graphene molecules [31], and stable radicals [32,33]. A detailed discussion concerning the choice of proper softwares for virtual FRP of vinyl monomers is presented elsewhere [3]. Digital twins of numerous various fullerene $C_{60}$ derivatives, which constitute the majority of the DTs' set in Table 1, were designed basing on the grounds of the fullerene $C_{60}$ spin chemistry [29,30]. Over thirty years ago, the final products of the relevant ERs were called fullerenyls [34]. The past decades since then, this name has taken root [35-38] and we will use it in what follows.

The obtained results, composing the wished IT product, concern equilibrium structural data of the studied DTs supplemented with thermodynamic and kinetic parameters of their formation. 'The same conditions' status of the obtained results implies the application of the same-type QC consideration as well as performing the ERs under study in vacuum at absolute temperature zero.

## 2. Potential FRCP of vinyls with stable radicals in light of suggested ERs and DTs. Digitalization instruments

In the chain-reaction concept, each polymerization process is the result of a severe competition between ERs following their rates. It is this principle that will be used below when determining the winner. However, any competition level is determined by the internal content of its participants and their number. Let us look at the list of Table 1 from this point of view. Reactions 1 and 2, uniting free radical with monomer and its oligomers, evidently govern the FRP of a monomer, the first digitalization of which is presented elsewhere [3]. Besides, reaction 1 determines the feasibility of any vinyl polymerization process as such.

**Table 1.** Nomination of elementary reactions and/or digital twins related to the initial stage of the free-radical copolymerization of vinyl monomers with stable radicals

| Reaction mark | Reaction equation [1] | Reaction rate constant | Reaction type |
|---|---|---|---|
| (1) | $R^\bullet + M \to RM^\bullet$ | $k_i$ | generation of monomer-radicals |
| (2) | $RM^\bullet + (n-1)M \to RM_n^\bullet$ | $k_p$ | generation of oligomer-radicals, polymer chain growth |
| (3a) | $F + M \to FM$ | $k_{2m}^F$ | two-dentant grafting of monomer on $C_{60}$ |
| (3b) | $F + M \to FM^\bullet$ | $k_{1m}^F$ | one-dentant stable radical grafting of monomer, generation of monomer-radical |
| (4) | $FM^\bullet + (n-1)M \to FM_n^\bullet$ | $k_p^F$ | generation of oligomer-radical anchored to $C_{60}$, polymer chain growth |
| (5) | $S + M \to SM^\bullet \equiv SM$ | $k_{1m}^S$ | one-dentant coupling with monomer |
| (6) | $F + RM^\bullet \to FRM$ | $k_{rm}^F$ | monomer-radical grafting on $C_{60}$ |
| (7) | $S + RM^\bullet \to SRM$ | $k_{rm}^S$ | monomer-radical capturing with stable radical |
| (8) | $F + R^\bullet \to FR$ | $k_R^F$ | free radical grafting on $C_{60}$ |
| (9) | $S + R^\bullet \to SR$ | $k_R^F$ | free radical capturing with stable radical |
| (10) | $F + S \to FS$ | $k_S^F$ | stable radical grafting on $C_{60}$ |
| (11) | $R^\bullet + FM^\bullet \to RFM$ | $k_{FM}^R$ | monomer-radical $FM^\bullet$ capturing with free radical |
| (12) | $S + FM^\bullet \to SFM$ | $k_{FM}^S$ | monomer-radical $FM^\bullet$ capturing with radical S |

[1] $M, R, F, S$ mark the objects' chemical attribution to vinyl monomer, initiating free radicals (either $AIBN^\bullet$ or $BP^\bullet$), stable radicals (fullerene $C_{60}$, and *TEMPO*), respectively. Superscript black spot distinguishes radical participants of the relevant reactions.

By selecting the most successful radical partner of this reaction empirically, the researchers opted for *one*-target free radicals such as $AIBN^\bullet$ and $BP^\bullet$. A similar stable radical *TEMPO* was found unsuitable for this role. As for the *multy*-target $C_{60}$, the first wave of polymer researchers, who introduced fullerene into polymerization and were confident in its radical nature, attempted repeatedly to detect the growth of a polymer chain anchored on the fullerene itself [8-20, 39]. However, no distinct evidence of the process has been found until now in spite of potential ability of such event was clearly vivid. The matter is that the intermolecular junction between $C_{60}$ and a vinyl monomer is configured with two $sp^2$C-C bonds, one belonging to fullerene and the other presenting a vinyl group of monomer thus dictating either two-dentant or one-dentant intermolecular junctions. This extreme fullerenes' potentiality has been largely discussed, for example, for fullerene dimers [40, 41]. As firstly shown recently for fullerene $C_{60}$ – vinyl monomer compositions using methyl methacrylate as an example [4], the first configuration provides the formation of [2x2]-cycloadded monoadduct fullerenyl $FM$, while the second leads to the generation of fullerene-grafted monomer-radical $FM^\bullet$. Equilibrated structures of these DTs in Figure 1 are supplemented with the atomic-chemical-susceptibility (ACS) distributions, revealing the radical properties of the bodies and expressed in the term of the unpaired-electron fractures $N_{DA}$ [29,30]. ACS plottings of both DTs are settled over the background one presenting the ACS distribution related to the initial fullerene $C_{60}$. As seen in the figure, both fullerenyls retain the *multi*-target type of the radicalization, similar to that of the patterned one. The appearance of new targets with increased radicality in the fullerene core (see the most prominent atoms 35, 22 and 11 in Figure 1a and extra atoms 37 and 35 in Figure 1b) is the expected consequence of the reconstruction of the fullerene $sp^2$C-C bond system caused by the derivative formation [29,30]. This characteristic feature of fullerenes underlies the ability of further polyderivatization of the molecules. However, the emergence of a new target ability of fullerenyl $FM^\bullet$ with a predominant $N_{DA}$ of 0.97 $e$ at atom 62, related to the vinyl bond of the monomer (see Figure 1b), exhibits the undeniable readiness of the latter to continue the association with other monomer molecules, similar to that is observed in the case of its FRP [3].

Obviously, this ability realization is governed with the thermodynamic and kinetic parameters of ER (3b). In the case of methyl methacrylate [4], reaction (3b) is not active enough to win the competition mentioned earlier so that the growth of polymer chains $FM^{\bullet}_{n+1}$ (ER (4)) attached to the fullerene core does not occur. In contrast to the above fullerenyls, $SM$ (ER (5)), when is formed, presents a routine non-radical one-bond-coupled intermolecular complex with a complete zeroing of all the $N_{DA}$ values.

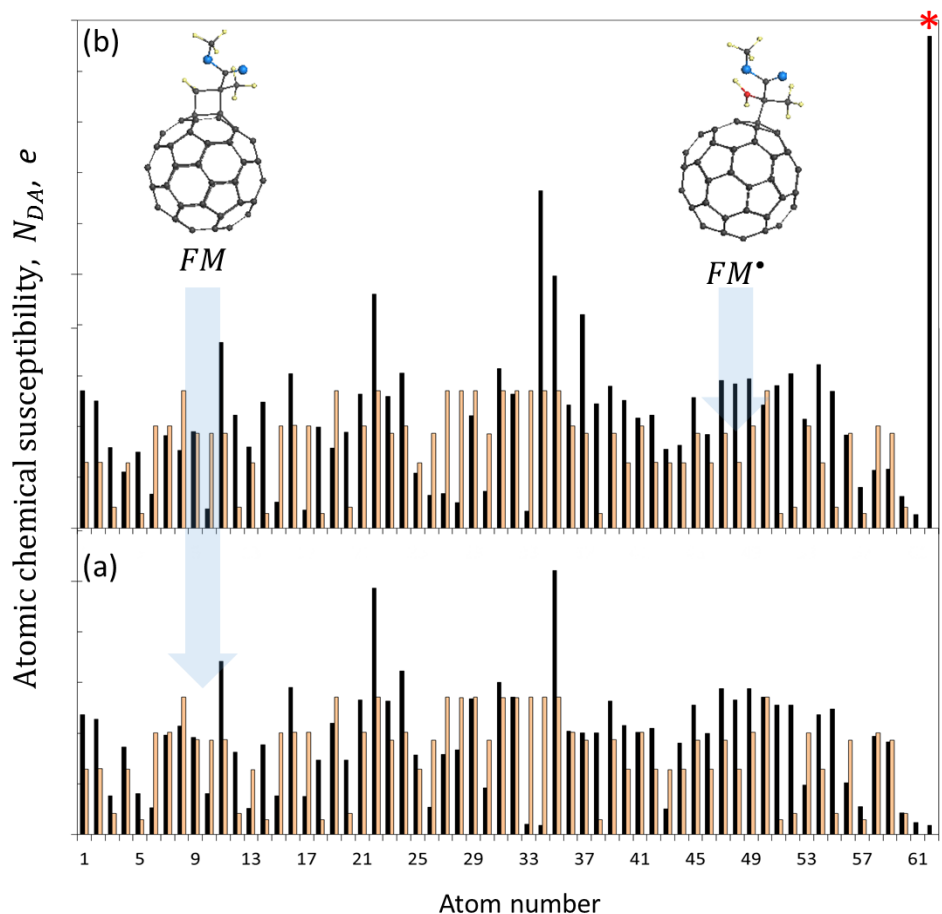

**Figure 1.** Equilibrium structures and ACS radicality of methyl-methacrylate fullerenyls $FM$ and $FM^{\bullet}$ (black histograms). Light rose marks the ACS radicality of fullerene $C_{60}$. The carbon-atom numeration of fullerene and fullerenyls is the same. UHF AM1 calculations.

In contrast to non-reactive monomer, the capturing of its monomer-radical, described by ERs (6) and (7), is traditionally highly expected for both stable radicals. Actually, these ERs are of particular importance having the opportunity to completely stop the polymerization process. Then follow ERs (8) and (9), revealing a similar capturing of free radicals $R^{\bullet}$. Both reactions evidently affect the monomer polymerization, decreasing the number of initiating free radicals. Reaction (10) takes into account the interaction of stable radicals between themselves, while reactions (11) and (12) describe the capturing of monomer-radical $FM^{\bullet}$ with stable ones. completes the ERs set and describes the interaction of stable radicals themselves. A detailed consideration of reactions (1) and (2) has been performed recently [3,4]. Reactions (3-12) are the main goal of the current study.

The thermodynamics and kinetics of the above ERs are considered on the basis of standard energy graphs, involving the total energy of the equilibrium reactant community $Y$, $E(Y)$, the energy of the product $X$ of the $Y$ pair interaction, $E(X)$, and the energy of the transition state of the molecular complex under consideration, $E_{TS}(X \leftrightarrow Y)$. There is also one more important energetic parameter – reaction enthalpy, $\Delta H$, or coupling energy, $E_{cpl}=E(X)-E(Y)$ (see Figure 2a). Energies $E(X)$ and $E(Y)$ are strictly evaluated via heats of formation of the final product and initial partners of the studied ER, respectively, while $E_{TS}(X \leftrightarrow Y)$ is determined as the maximum of the barrier dissociation profile presented in Figure 2b for the styrene + $AIBN^\bullet$ pair presenting one of the ER of type 1 (see the detailed description of the profile procedure in [3,4]).

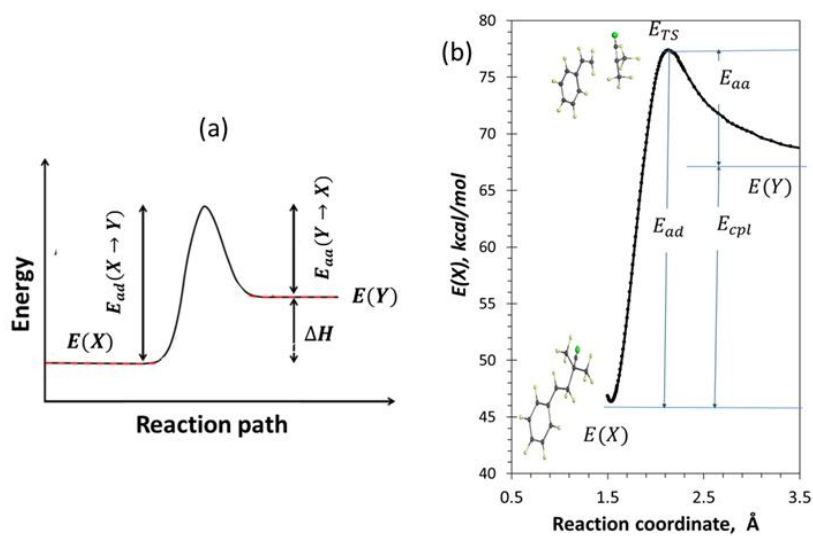

**Figure 2**. Standard energy graph (a) and barrier profiles of the dissociation of the monomer-radicals $RM^\bullet$ composed of styrene and free radical $AIBN^\bullet$ (b). UHF AM1 calculations.

As for the reaction kinetics, its standard description concerns the rate constant, $k(T)$, which is expressed through the Arrhenius equation [40-43]

$$k(T) = Ae^{\left(\frac{-E_a}{kT}\right)}. \qquad (1)$$

Here $A$ is a complex frequency factor, while $E_a$ presents the activation energy, which is either the energy of the $X$ product dissociation, $E_{ad}$, or the $Y$ pair association, $E_{aa}$. Since the frequency factor for one-type elementary reactions is expected to change weakly [5, 40-43], so that activation energy becomes governing. Its value can be determined by building barrier profiles of either association or dissociation of molecular pairs $Y$ and $X$, respectively. As seen in Figure 2b, both values can be strictly determined when examining the barrier profiles. The latters consist the basic pool for the comparative analysis of the kinetic efficiency of different ERs.

Joint comparable consideration of a set of ERs, digitized in a unified way, is a distinctive feature of the DT concept. As has been mentioned earlier, this approach allows to confidently identify trends inherent in the object or process under consideration. The current study also did not go beyond the scope of this practice and made it possible to identify the main structural and kinetic features of polymerization systems that control these trends.

## 3. Results

### 3.1. Introductory comments

Simultaneous consideration of the ten ERs presented above imposes certain conditions on the manner of joint presentation of the results. It turned out that representing the latter in the form of a matrix, the framing boundaries of which are marked by the participants in the polymerization process under consideration, is the most appropriate. As first used in the case of methyl methacrylate [4], Table 2 presents such a combined set of data relating to the polymerization of styrene in the presence of the initiating free radical $AIBN^\bullet$ and small additions of the stable radicals TEMPO and $C_{60}$. In certain way, matrix Table 2 is a visualization of a 'pineapple-on-the-plantation' essence of the polymerization [3] in the realities of the FRP of styrene and its copolymerization (FRCP) with two stable radicals. The table consists of three parts. The first part lists nominations, which mark both the considered ERs and their final products simultaneously. The latters alongside with the headings constitute a set of the considered DTs. The second part involves thermodynamic descriptors related to the ERs in terms of coupling energy $E_{cpl}$ of the relevant final products. The third part of the table concerns the kinetic descriptors in term of activation energy related to the dissociation of these products $E_a \equiv E_{ad}$.

**Table 2.** Elementary reactions and their final products supplemented with virtual thermodynamic and kinetic descriptors related to the FRCP of styrene with stable radicals *TEMPO* and $C_{60}$, while initiated with $AIBN^\bullet$ free radical

| | $M$ | $R^A M^\bullet$ | $AIBN^\bullet$ ($R^{A\bullet}$) | $TEMPO$ ($S^\bullet$) | $C_{60}$ ($F$) 2-dentate | $C_{60}$ ($F$) 1-dentate |
|---|---|---|---|---|---|---|
| | | | **Digital Twins' set** | | | |
| | | | | | FM | $FM^\bullet$ |
| $M$ | $M_{n+1}$ | $R^A M^{\bullet\bullet}_{n+1}$ | $R^A M^\bullet$ | $SM$ | FM | $FM^\bullet$ |
| $R^A M^\bullet$ | - | $(R^A M)_2$ | $R^A R^A M$ | $SR^A M$ | $FR^A M$ | |
| $R^{A\bullet}$ | - | - | - | $SR^A$ | $FR^A$ | |
| $S^\bullet$ | - | - | - | - | $FS$ | |
| $FM^\bullet$ | $FM^{\bullet\bullet}_{n+1}$ | - | $R^A FM$ | $SFM$ | - | |
| | | | **Coupling energies, $E_{cpl}$, kcal/mol** [1] | | | |
| $M$ | | -7.38 ÷ -29.01 (2-6) | $-22.51$ (1) | 4.58 | -34.31 | -18.59 (1) |
| $R^A M^\bullet$ | - | | -29.80 | -5.03 | -19.11 | |
| $R^{A\bullet}$ | - | - | - | 5.01 | -20.14 | |
| $S^\bullet$ | - | - | - | - | 0.98 | |
| $FM^\bullet$ | -13.86 (2) -18.81 (3) -18.88 (4) -23.04 (5) | - | -19.63 | -0.89 | - | |
| | | | **Activation energies, $E_a$, kcal/mol** [1,2] | | | |
| $M$ | | 6.12 - 16.49 [3] (2-6) | 8.50 (1) | 19.28 | 24.52 | 8.38 (1) |
| $R^A M^\bullet$ | - | - | not defined | 19.04 | 9.73 | |
| $R^{A\bullet}$ | - | - | - | 18.74 | 9.41 | |
| $S^\bullet$ | - | - | - | - | not grafted | |
| $FM^\bullet$ | 11.25 (2) | - | not defined | 19.54 | - | |

[1] Digits in brackets mark the number of monomers in the oligomer chain.
[2] Bold data are determined from the barrier profiles similar to shown in Figure 2b.
[3] The data are calculated by using Evans-Polany-Semenov relation presented in [3].

'Matrix elements' of the table are evidently divided into four groups marked with different colors. Yellow elements present ERs alongside with the relevant DTs that govern FRP of styrene. Elements in light blue describe ERs and DTs related to the FRCP of styrene with *TEMPO*, while faint-pink elements do the same with respect to the FRCP of styrene with $C_{60}$. ER (4) and the related DTs concerning the fullerene-stimulated polymerization of styrene are highlighted in dark pink.

### 3.2. Virtual participants of the polymerization process

Equilibrium structures of the DTs nominated in Table 2 are presented in Figure 3, thus visualizing a complexity of the considered polymerizable medium following the row arrangement of nominated DTs in the table. Thus, the table headings are represented in the figure by image-set (a). The set includes the main initial participants of the polarization process usually considered in the case of FRP - a monomer and four radicals. The first three of the latters are one-target, reflecting the fact that the reactivity of each of them is concentrated on a single atom. These atoms, carbon in the case of $R^A M^\bullet$ and $R^{A\bullet}$ as well as oxygen in $S^\bullet$, are highlighted in red. The fourth member of this chain is occupied by a fullerene $C_{60}$, radical efficiency of which is distributed over 60 atoms of the molecule, remaining the same within each of five selected groups of atoms but varying between the groups, which is highlighted in different colors in the insert [29,30,42]. This color representation evidences the multitargetability of the molecule, whose 12 light gray atoms are the most active once characterized by the largest $N_{DA}$ value.

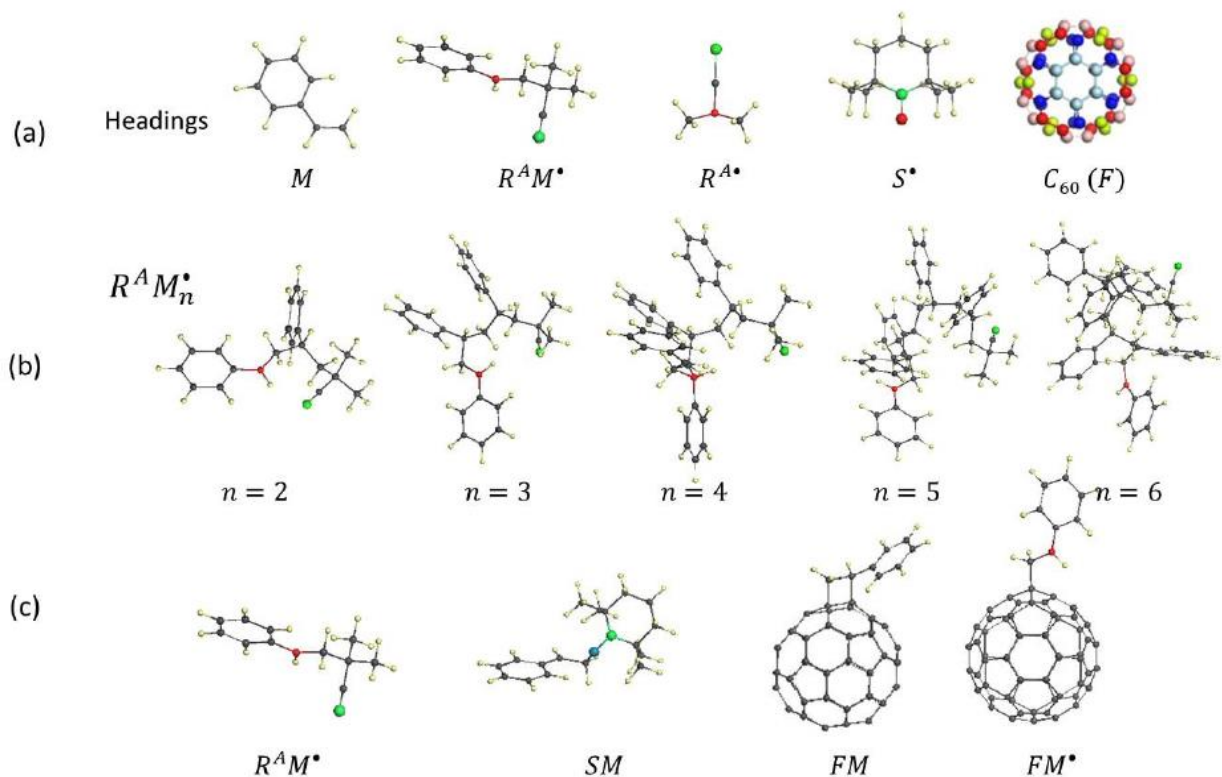

**Figure 3.** Equilibrium structures of digital twins related to the FRCP of styrene with stable radicals. (a) Structures, representing headings of Table 2. Colored image of $C_{60}$ exhibits a specific spin density distribution over the molecule carbon atoms [29,30,42]. (b) Oligomer radicals $R^A M^\bullet_{n+1}$ for $n$ from 2 till 6. (c) Digital twins produced by intermolecular interaction of styrene with free radical RA, TEMPO and $C_{60}$, both two-dentant and one-dentant trapping. UHF AM1 calculations.

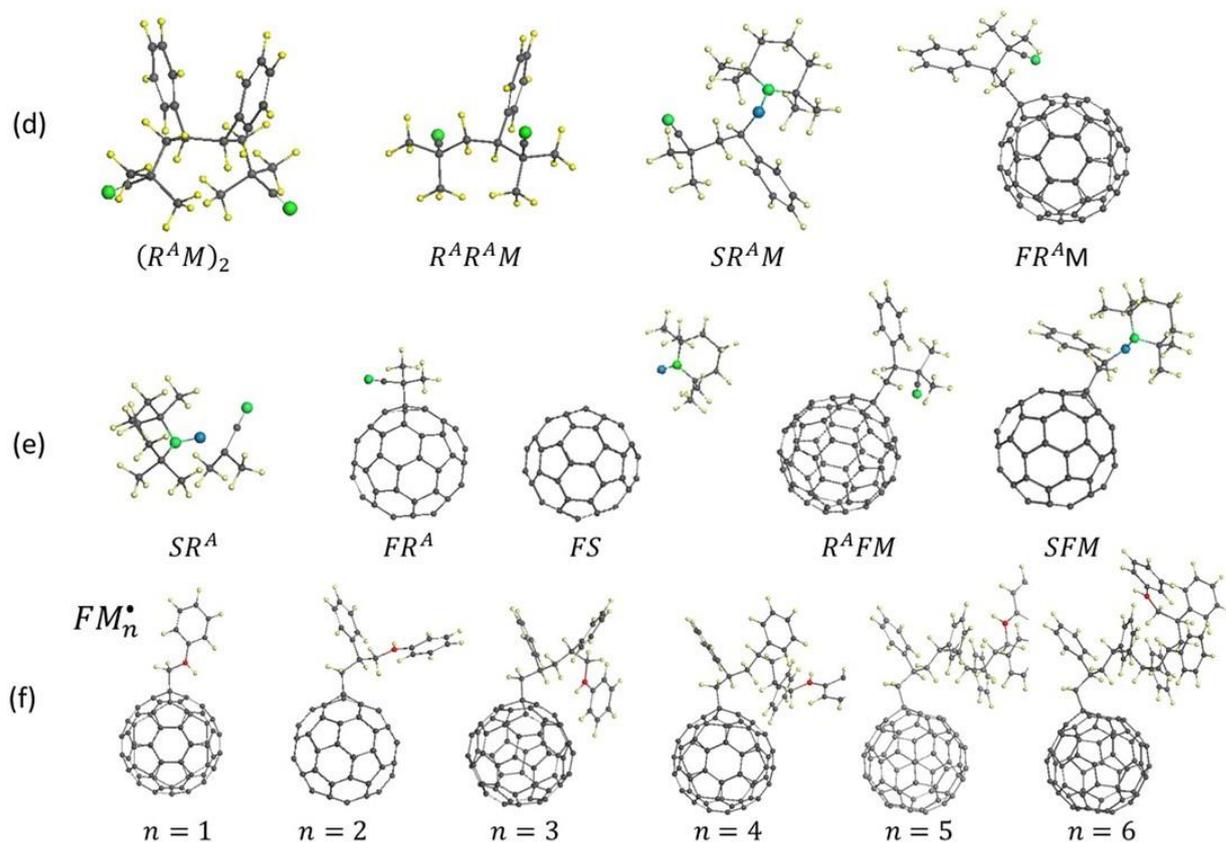

**Figure 3** cntd. (d) DTs describing the interaction of monomer-radical $R^A M^\bullet$ with itself, free radical $R^{A\bullet}$, TEMPO, and fullerene $C_{60}$, respectively; (e) DTs related to the interaction of radicals between themselves - $R^{A\bullet}$ with TEMPO and fullerene, fullerene with TEMPO, $R^{A\bullet}$ and TEMPO with $FM^\bullet$; (f) Oligomer radicals $F(M)_{n+1}^\bullet$ for $n$ from 1 till 6. Small yellow and gray balls mark hydrogen and carbon atoms, respectively. Larger green and blue balls depict nitrogen and oxygen atoms. Red balls mark carbon (small) and oxygen (large) target atoms. UHF AM1 calculations

Digital twins of the first row of Table 2 are exhibited in Figure 3 by image-sets (b) and (c). The former represents a growth of the styrene polymer chain via a continuous length increasing of the styrene oligomer-radicals $R^A M_{n+1}^\bullet$, described in details earlier [3]. The next set unites the remaining four matrix elements of the row, two of which, namely $SM$ and $FM$, are standard stable intermolecular complexes, characterizing with zeroth $N_{DA}$ on all atoms, while $R^A M^\bullet$ and $FM^\bullet$ are radicals with the strongest concentration of radicality of $N_{DA}$ = 0.67 $e$ on carbon atoms of the styrene vinyl group in both cases. This similarity of the two bodies was expectedly manifested later in the polymerization of styrene in both cases.

Four DTs of the table's second row form the image-set (d) in the figure. The series presents all the cases of potential capturing of monomer-radical $R^A M^\bullet$, which leads to the termination of the polymer chain $R^A M_{n+1}^\bullet$ growth. Two first of the series, considered in details earlier [3], can be attributed to self-inhibiting. However, the very fact of successful and almost instantaneous empirical FRP of styrene indicates the negligible role of these participants in the polymerization process, so they will not be considered further. Single DTs remaining in the third and fourth rows of the table form image-set (e). And, finally, a series (f) of images $FM_{n+1}^\bullet$, located in the last row of the table, completes the structural-atomic representation of the considered DTs in Figure 3.

As can be seen from the above, visualization of DTs significantly enlivens the polymerizable medium and makes its analysis more presentive. At the same time, it becomes the best way to trace the polymerization process when replacing any participant in this process by another one as well as to suggest the first predictions. Thus, a comparison of previously published results [4] with the current work reveals highly important role of monomer, thus establishing that the FRCP of methyl methacrylate and styrene with stable radicals occurs differently. As for the role of free radicals, Table 3 and Figure 4 exhibit changes that occur in both FRP and FRCP of styrene when the initiating alkyl-nitrile radical $AIBN^{\bullet}$ is replaced by benzoyl-peroxide $BP^{\bullet}$. Figure 4 presents DTs, only which from the total content of Table 3 react on the free radical replacement. As seen in the figure, the first innovation concerns the monomer-radical $R^P M^{\bullet}$ itself and the resulting polymer chain $R^P M^{\bullet}_{n+1}$. This first obtained virtual chain, represented by the set (b) in Figure 4 and consisting of monomer-radical and two oligomer ones, generally has much in common with that one previously initiated by the $AIBN^{\bullet}$ [3]. Naturally, the structural details of the chain, determined by intermolecular junctions as well as by changes in these junctions composition in space, are quite different. This circumstance, however, does not at all affect the methodology for digitalizing the properties of such polymer chains that was proposed earlier [3]. Finally, series (c) in Figure 4 presents results of capturing of either monomer-radical $R^P M^{\bullet}$ or free radical $R^{P\bullet}$ with stable radicals *TEMPO* and $C_{60}$.

**Table 3.** Elementary reactions and their final products supplemented with virtual thermodynamic and kinetic descriptors related to the FRCP of styrene with stable radicals *TEMPO* and $C_{60}$, while initiated with $BP^{\bullet}$ free radical

| | $M$ | $R^P M^{\bullet}$ | $BP^{\bullet}$ ($R^{P\bullet}$) | *TEMPO* ($S^{\bullet}$) | $C_{60}$ (F) 2-dentate | $C_{60}$ (F) 1-dentate |
|---|---|---|---|---|---|---|
| **Digital Twins' set** | | | | | | |
| $M$ | $M_{n+1}$ | $R^P M^{\bullet\bullet}_{n+1}$ | $R^P M^{\bullet}$ | $SM^{\bullet}$ | $FM$ | $FM^{\bullet}$ |
| $R^P M^{\bullet}$ | - | $(R^P M)_2$ | $R^P R^P M$ | $SR^P M$ | $FR^P M$ | |
| $R^{P\bullet}$ | - | - | | $SR^P$ | $FR^P$ | |
| $S^{\bullet}$ | - | - | - | - | $FS$ | |
| $FM^{\bullet}$ | $FM^{\bullet\bullet}_{n+1}$ | - | $R^P FM$ | $SFM$ | | |
| **Coupling energies, $E_{cpl}$, kcal/mol [1]** | | | | | | |
| $M$ | | -23.23 (2) <br> -18.47 (3) | -40.33 (1) | 4.58 | -34.31 | -18.59 (1) |
| $R^P M^{\bullet}$ | | | | -2.85 | --23.12 | |
| $R^{P\bullet}$ | | | | --0.63 | -40.62 | |
| $S^{\bullet}$ | | | | - | 0.98 | |
| $FM^{\bullet}$ | -13.862 (2) <br> -18.813 (3) <br> -18,879 (4) <br> -23.042 (5) | - | -19.63 | -0.89 | | |
| **Activation energies, $E_a$, kcal/mol [1,2]** | | | | | | |
| $M$ | | 8.78 (2) | 2.79 (1) | 19.28 | 24.52 | 8.38 (1) |
| $R^P M^{\bullet}$ | | | | 17.60 | 7.02 | |
| $R^{P\bullet}$ | | | | not coupled | 28.82 | |
| $S^{\bullet}$ | | | | - | not grafted | |
| $FM^{\bullet}$ | 11.25 (2) | - | not determined | 19.54 | | |

[1] Digits in brackets mark the number of monomers in the oligomer chain.
[2] Bold data are determined from the barrier profiles similar to shown in Figure 2b.

Analogous passport-like table-figure pairs can be recommended to be created in any case of both empirical and virtual study of the FRP and/or FRCP of vinyl (as well of allyl and any other) monomers to make the scene of polymer games more vivid and clear.

### 3.3. Grounds of virtual thermodynamics and kinetics of the polymerization

Second parts of Tables 2 and 3 present thermodynamic descriptors of the considered ERs in terms of the reaction energies, $\Delta H$, or coupling energies, $E_{cpl}$, related to their final products. The $E_{cpl}$ values are counted from the zero energy level, set equal to $E(Y)$. Analyzing these quantities, let us pay attention to a strong connection between them and the activation energies $E_{cpl}$ of any ER. As follows from Figure 2a, negative values of $E_{cpl}$ mean that $E_{ad} > E_{aa}$ indicating that the association of the initial reagents $Y$ and the formation of a stable final product $X$ is kinetically more justified than the decomposition (instability) of the latter. If the $E_{cpl}$ values are positive, then the situation is reverse and decomposition of the final product is more likely.

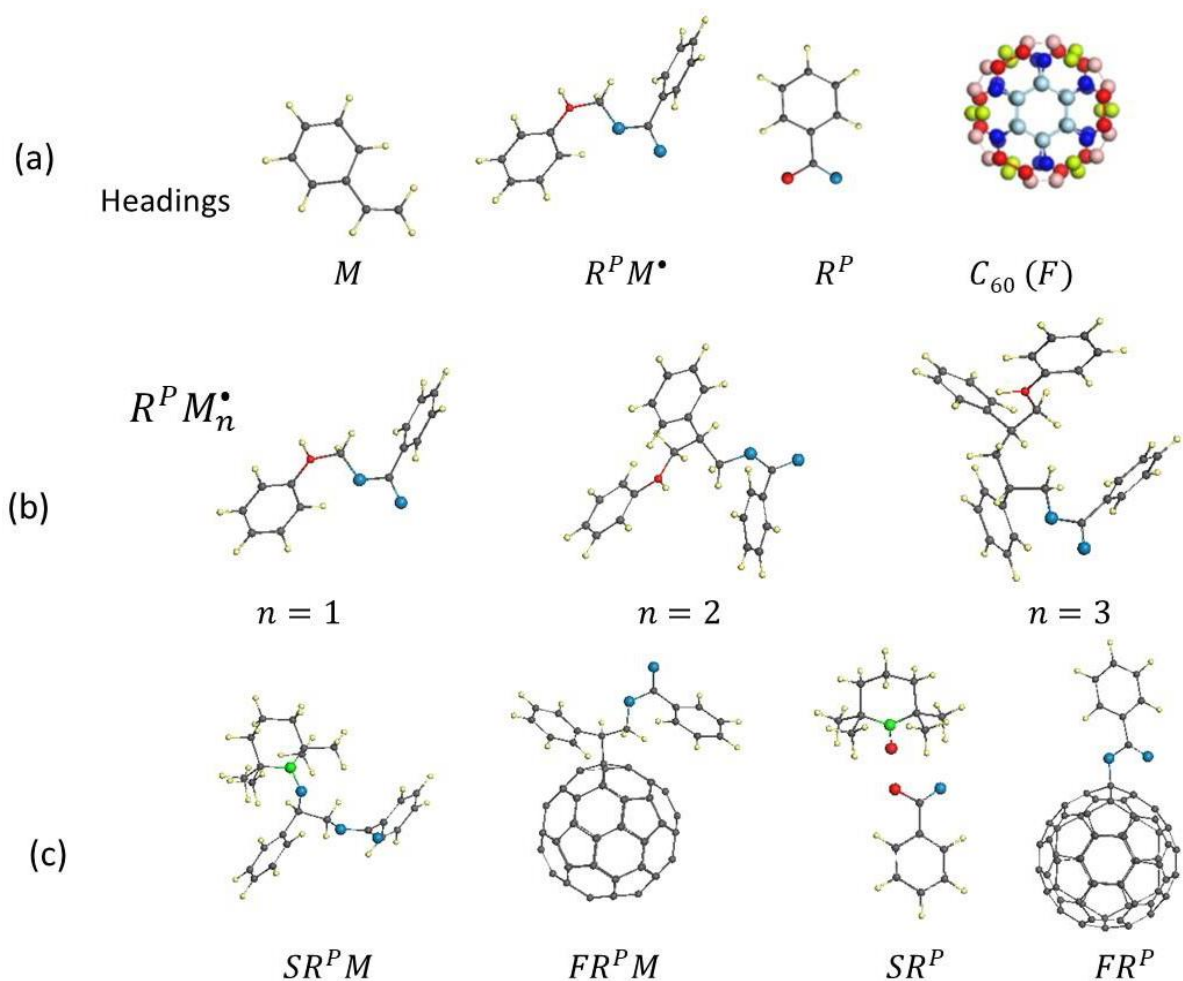

**Figure 4.** Equilibrium structures of digital twins of FRCP of styrene with stable radicals. (a) Structures, representing headings of Table 3. Colored image of $C_{60}$ exhibits a specific spin density distribution over the molecule carbon atoms [29,30,42]. (b) Oligomer radicals $R^P M^{\bullet}_{n+1}$ for $n$ from 1 till 3. (c) Digital twins produced by intermolecular interaction of styrene monomer-radical $R^P M^{\bullet}$ with *TEMPO* and $C_{60}$, as well as free radical $R^P$ with *TEMPO* and fullerene. Small yellow and gray balls mark hydrogen and carbon atoms, respectively. Larger green and blue balls depict nitrogen and oxygen atoms. Red balls mark carbon (small) and oxygen (large) target atoms. UHF AM1 calculations

As can be seen from Tables 2 and 3, most of the associative reactions listed there are kinetically possible, so the polymerization process is, indeed, a competition of a large number of participants, in which the win goes to the fastest.

In the current study, the kinetic efficiency of ERs is discussed in terms of $E_{aa} \equiv E_a$ values. Having at hand a large set of $E_a$ values and grouping them in accordance with groups of one-type reactions, one could hope for the applicability of a simplified method of determining activation energies, using the Evans-Polyani-Semenov (EPS) rule [5-7]. The latter is based on the assumed linear relationship between $E_{cpl}$ and $E_a$ values. However, as it turned out, this rule works for some ERs only, such as, for example, $R^A M_n^\bullet$ and $R^P M_n^\bullet$, while it does not work at all between DTs related to different ERs. Therefore, to determine the $E_a$ values reliably, the method of virtual decomposition barriers related to final products of the considered ERs was chosen. The obtained data are accumulated in the third parts of Tables 2 and 3.

Figures 5 and 6 present selected examples of standard and complicated situations met under the way of the digitalizing of various barrier profiles. A typical situation, when all the features of the energy graphs $E(R)$ are well defined, is shown in Figure 5a. Two main extremes of the plottings attributed to their minima and maxima are clearly vivid, thus providing a direct evaluation of $E_{ad}$ and $E_{aa}$ values. The examples given relate to the first two members of the styrene polymer chain $FM_{n+1}^\bullet$ ($FM^\bullet$ and $FM_2^\bullet$) attached to the fullerene and to the $FM$ describing the styrene two-dentant attaching to the fullerene. Noteworthy are the well-defined maxima of the graphs, which record the energy of the transition state $E_{TS}$ of the products of the corresponding ERs. In all cases, the intermolecular junctions in the products, which play the role of reaction coordinate, are determined by an $sp^3$C-CH bond, which is broken in the transition state [3]. The latter was evidenced by the fact that the position of the energy graphs maxima coincides with $R_{crit}^{C-C}$ of 2.11±0.1 Å that determines maximum length of the sp³C-C bond, above which the bond becomes radicalized thus revealing the start of its breaking [43,44]. Oppositely to the case, the sp³C-O bond forms the junctions in $SFM$ and $SR^AM$ products presented in Figure 5b. Expectedly, the maximum positions of their barrier profiles should differ from those provided with breaking the sp³C-C bond that is a reaction coordinate of the $FR^AM$ decomposition. Actually, these positions constitute 1.91 and 2.21 Å, which well correlates with $R_{crit}^{C-O}$ of 2.01 Å relating to the dissociation of the sp³C-O bond of ethylene glycol presented in the bottom of the figure. Evidently, $R_{crit}^{C-O}$, as well as $R_{crit}^{C-C}$, deviates in different atomic surrounding, which was observed in the current study. Shown in Figure 5c is related to the case, when the determination of $E_{TS}$ becomes uncertain. Two DTs presented in the figure have the same atomic compositions while formed (and reverse decomposed) differently. The first DT $FR^AM$ is the result of capturing monomer-radical $R^AM^•$ with fullerene. Accordingly, the intermolecular junction is formed by the sp³C-C bond, marked with red, which is the reaction coordinate when the product is decomposed. In contrast, DT $R^AFM$ presents the case when monomer-radical $FM^•$ is inhibited with free radical $R^{A•}$. The intermolecular junction as well as the reaction coordinate is transmitted to the sp³C-O bond. The corresponding bonds are red-marked. As seen in the figure, if the determination of the corresponding energy $E_{TS}$ is not difficult in the first case, in the second case it becomes uncertain.

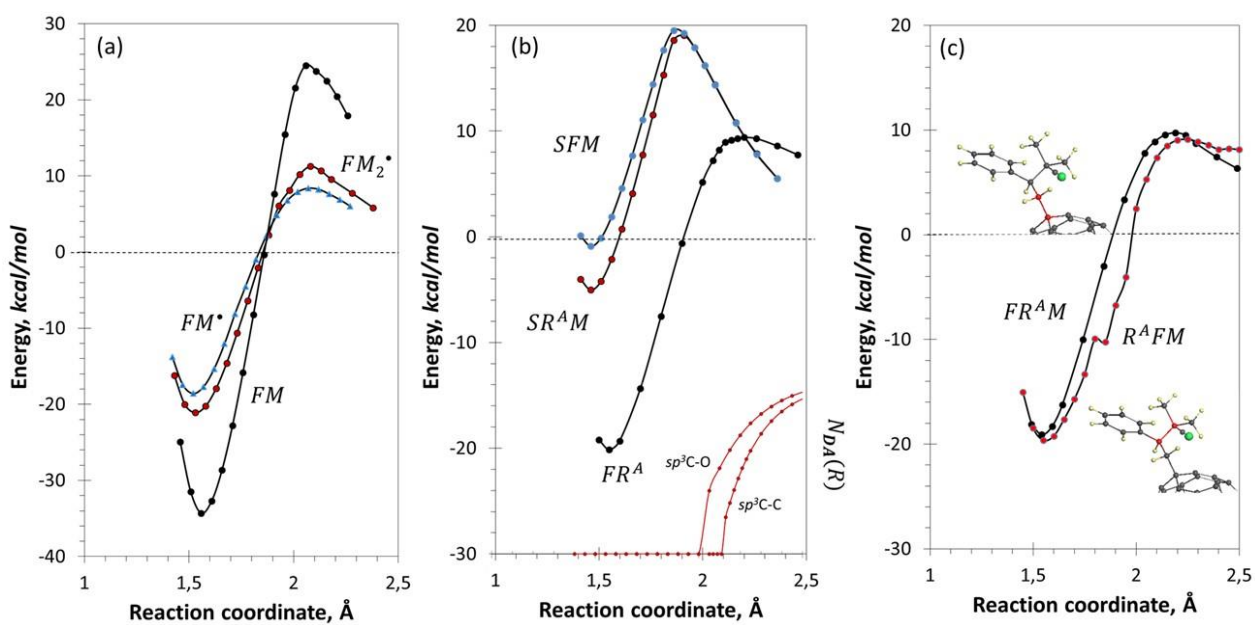

**Figure 5.** Virtual barrier profiles of the DTs decomposition. DTs are the following: (a) two first members of the styrene polymerization $FM_{n+1}^\bullet$ ($FM^\bullet$ and $FM_2^\bullet$) and $FM$ (two-dentant attachment of styrene to the fullerene); (b) $FM^\bullet$ and $R^A M^\bullet$ monomer-radicals captured with *TEMPO* (*SFM* and $SR^A M$) as well as $R^{A\bullet}$ radical captured with fullerene ($FR^A$); (c) $R^A M^\bullet$ monomer-radical captured with *TEMPO* ($SR^A M$) while $FM^\bullet$ one with free radical $R^{A\bullet}$ ($R^A FM$). $N_{DA}(R)$ graphs are related to the elongation of the $sp^3$C-C bond of ethane and $sp^3$C-O bond of ethylene glycol [44]. UHF AM1 calculations.

A particular behavior of the sp³C-O and its inability to play the role of the reaction coordinate is justified by the players of the styrene FRP game when the initiating free radical $AIBN^•$ is replaced with $BP^•$. Figure 6 presents two pairs of DTs where this coordinate of one partner (either $R^PM_n^•$ (a) or $FR^PM$ (b)) is played with the sp³C-C bond, while the sp³C-O bond plays the role for other partners ($R^PM^•$ (a) and $R^PFM$ (b)). As seen in the figure, in the first case the $E_{TS}$ values are well evaluated and localized within the region of the expected values for $R_{crit}^{C-C}$. In the second case, this evaluation becomes uncertain. Evidently, the radical character of the considering ERs is the feature's reason. However, more detailed answering the question requires further investigation.

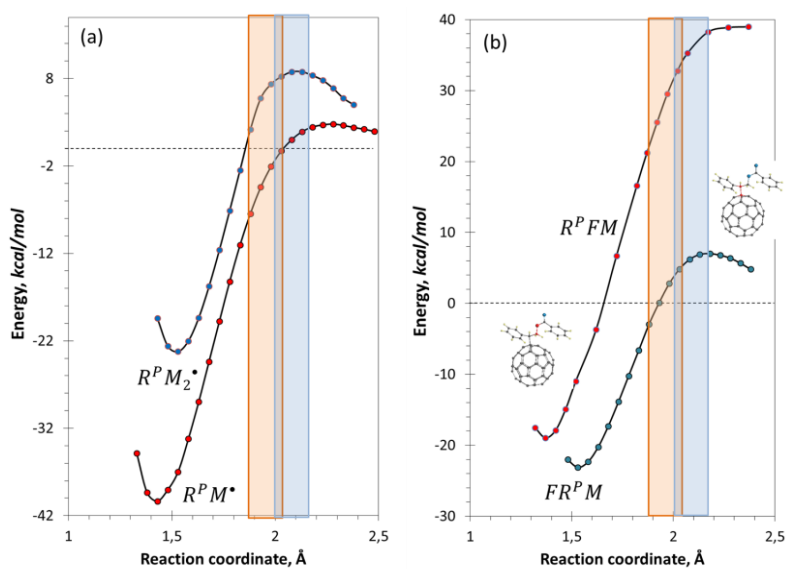

**Figure 6.** Virtual barrier profiles of the DTs decomposition. DTs are related to the following ERs: (a) the formation of monomer- and dimer-radicals $R^PM^•$ and $R^PM_2^•$, respectively; (b) capturing of $R^PM^•$ monomer-radical with fullerene ($FR^PM$) while of $FM^•$ one with free radical $R^{P•}$ ($R^PFM$). Light pink and light blue bands mark assumed deviation intervals of critic values of the maximum length of the sp³C-C and sp³C-O bonds under breaking. UHF AM1 calculations.

## 4. Virtual and real kinetics of the polymerization process under consideration

Figure 7 accumulates empirical data related to kinetics of the initial stage of the FRP of styrene and its FRCP with fullerene $C_{60}$ and TEMPO, while being initiated with either $AIBN^\bullet$ or $BP^\bullet$ free radicals. The panorama presents a complex of experiments performed under the same conditions concerning the temperature, solvent, contents of monomer, free radicals and stable radicals [22,23]. Curves 1 in all the panels present FRP of styrene that are referent. Figure 7a exhibit the effect of small additives of $C_{60}$ on the referent process. Figure 7b shows effect of combined action of $C_{60}$ and TEMPO, Figure 7c does the same concerning the replacement of $AIBN^\bullet$ with $BP^\bullet$ followed with small additive of $C_{60}$. The observed effects are well pronounced and strong. Virtual $E_a$ data listed in Tables 2 and 3 do not pretend on the reproduction of the empirical monomer-conversion dependences discussed above, but they allow us to hope for identifying the key reasons for their behavior.

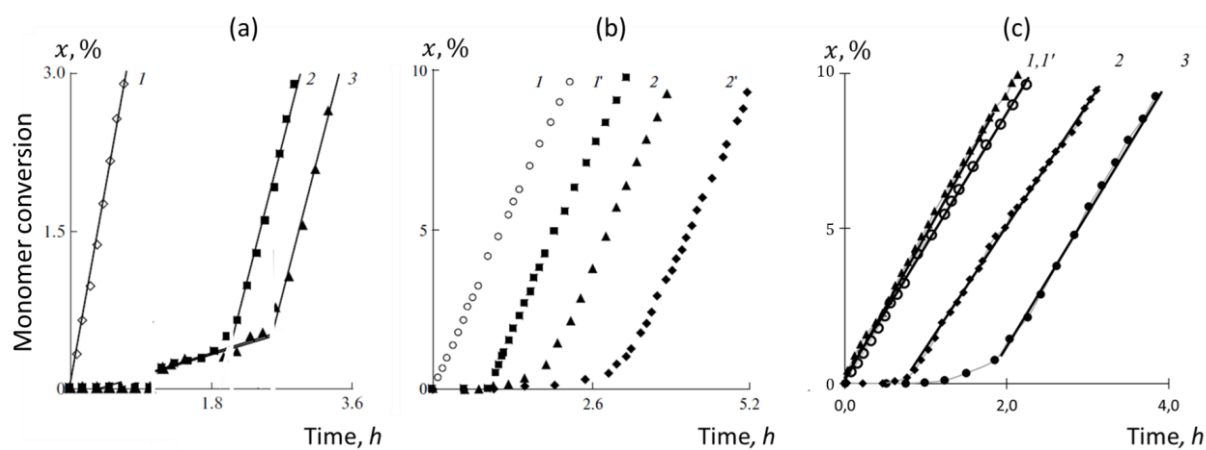

**Figure 7.** Empirical kinetics of the initial stage of both FRP of styrene and its FRCP with *TEMPO* and fullerene $C_{60}$. $AIBN^\bullet$-initiated conversion of styrene in the presence of different [$C_{60}$] 0 (curve 1); $1.0 \times 10^{-3}$ (curve 2); $2.0 \times 10^{-3}$ mol/L (curve 3) (a); (*1*) - [*TEMPO* ($C_{60}$)] 0; (*1'*) - [*TEMPO*] $1.0 \times 10^{-3}$ mol/L; (*2*) - [$C_{60}$] $1.0 \times 10^{-3}$ mol/L; (*2'*) - [*TEMPO* ($C_{60}$)] $1.0 \times 10^{-3}$ mol/L (b). (c) Conversion of styrene, initiated with $AIBN^\bullet$ (1' and 3) and $BP^\bullet$ (1 and 2) (c) in the absence (1' and 1) and in the presence (3, 2) of fullerene [$C_{60}$]=2.0 mol/L. T=60⁰C; o-DCB solvent; [St]=2.0 mol/L; [AIBN (BP)]=$2.0 \times 10^{-2}$ mol/L. Digitalized data of Refs. 22 and 23.

Let us start with Figure 7a. As stated by chemists [22,23], the observed empirical effect concerns mainly the presence of the induction period (IP) caused by the fullerene addition. Therewith, the IP elongates with increasing fullerene content, and the conversion rate of the monomer massive polymerization, occurred after the IP termination, remains practically unchanged. Taken together these features were interpreted by the inhibition action of fullerene followed with its full consumption occurred by the IP end. Neither type of the action nor clearly observed slope of the IP straight line were interpreted unambiguously.

Applying to the virtual data, the figure content is related to those presented in yellow and faint pink in Table 2. The former are related to the FRP of styrene, while the latter – to the FRCP of styrene with $C_{60}$. As seen from the table, the FRP is governed with $E_a$ data filling the range from 6.1 to 16.5 kcal/mol. The values are typical for vinyl monomers [3] and are kinetically quite favorable for their FRP to be successful. The corresponding ERs include the initiation of the monomer-radical $R^AM^\bullet$ and the successive growth of the polymer chain $R^AM^\bullet_{n+1}$. As for the effect of $C_{60}$, three ERs, involving the formation of the $FM^\bullet$ monomer-radical and two stable fullerenyl radicals $FR^AM$ and $FR^A$, play the main role. Generally, listed $E_a$ data are not enough to judge which ER is the fastest, so that frequency factors $A$ evidently are needed. However, in the case of similar ERs, which is the case of ERs 3b, 6, and 8 (see Table 1) that are related to styrene FRCP with $C_{60}$, $E_a$ values can be used as a basis for the interrelation between the latters to be suggested. The following reaction-constant series looks like

$$k_{1m}^F > k_R^F > k_{rm}^F, \qquad (2)$$

indicating that polymerization of styrene on fullerene is the fastest.

The relation $k_{1m}^F > k_R^F$ is supported empirically when studying the $C_{60}$ conversion in the presence of either both styrene and $AIBN$ or $AIBN$ only, thus exhibiting a fivefold decreasing of the rate in the second case, which reveals a drastic difference in $FR^A$ and $FR^AM$ ERs. Moreover, empirically observed rate of the $AIBN$ solution remains unchanged when not styrene, but methyl methacrylate is input [23]. This fact is explained in a recent study of the virtual kinetics of methyl methacrylate, which established the dominant role of $FR^A$ reaction in its FRCP with fullerene [4]. This feature is evidenced with a remarkable decreasing of the referent monomer-conversion curve slope that becomes less and less when the fullerene concentration increases [22]. Therewith, no traces of the IP presence is fixed. Therefore, in contrast to methyl methacrylate case, as follows from Figure 7a, $FR^A$ reaction does not govern the FRCP in the styrene one.

Thus, the reasons of the empirically observed IP in Figure 7a and, consequently, the fate of the styrene polymerization on fullerene are decided with the relation between the rate constants $k_{1m}^F$ and $k_i$ for ERs (3b) $FM^{\bullet}$ and (2) $R^AM^{\bullet}$, respectively. As follows from Table 2, $E_a$-based $k_{1m}^F$ only slightly exceeds $k_i$. However, pre-exponential frequency factor $A$ is reasonably expected to be bigger for fullerenyl $FM^{\bullet}$ than that of much less monomer-radical $R^AM^{\bullet}$, thus supplying the required inequality $k_{1m}^F > k_i$. Accordingly, we have to conclude that the input of fullerene into a chemical reactor, intended for FRP of styrene, cancels $R^AM^{\bullet}$ reaction, replacing it with $FM^{\bullet}$one, thus starting polymerization of the monomer driven by the monomer-radical $FM^{\bullet}$. This explains a drastic consumption of $C_{60}$ in the styrene-load reactor during the initial stage of the polymerization [23]. This allows concluding that the reaction $FM_{n+1}^{\bullet}$ governs this stage because of which for styrene monomers

$$k_{1m}^F > k_R^F. \qquad (3)$$

Naturally, the three orders of magnitude difference in the contents of styrene and $C_{60}$ practically suppresses the $FM_{n+1}^{\bullet}$ conversion dependence against the background of that for $R^AM_{n+1}^{\bullet}$ to almost zero, which makes its appearance in the overall picture of monomer conversion similar to close-to-zero IP, which we see in Figure 7a. As seen, the monotonically increasing conversion initially occurs spasmodically. This behavior, however, is within the experimental error. Curve 2, repeated later in the next experiments, presented in Figures 7b and 7c (curves 2 and 3, respectively), has a well-developed expected cross-sectional appearance of a rounded hockey stick. Therefore, the FRCP of styrene with $C_{60}$ starts with the polymerization of styrene, both stimulated with and anchored at fullerene. When all fullerene molecules are consumed, a FRP of styrene, initiated with a free radical $AIBN^{\bullet}$, proceeds.

Let us move on to the situation when *TEMPO* is input to the reaction medium, involving styrene and $AIBN$. As can be seen from Figure 7b (curves 1 and 1'), the addition of *TEMPO* is accompanied by the appearance of a classical IP with an almost zero slope of its curve to the coordinate axis. Upon reaching the end of this period, the FRP styrene begins, whose conversion is practically identical to the reference one. In Table 2, ERs concerning TEMPO are in the cells marked in light blue. As follows from the table, three ECs, namely, $SM$, $SR^AM$ and $SR^A$ are possible, markedly differing thermodynamically. Two of them ($SM$ and $SR^A$) are characterized by positive $E_{cpl}$ pointing to the relation $E_{da} > E_{aa}$. As discussed previously, it is difficult to imagine that these reactions can be effective in a complex polymerization medium, full of competing participants. In contrast to the above two, ER $SR^AM$ is evidently thermodynamically favorable. Under these conditions, a choice in favor of the $SR^AM$ reaction is made taking into account both $E_a$ and $E_{cpl}$ values. This reaction corresponds to the capturing of monomer-radicals $R^AM^\bullet$ with *TEMPO*, terminates the FRP of styrene and confidently explains the presence of IP on the monomer-conversion curve with a zero slope relative to the abscissa.

Nevertheless, a question remains, why reaction $SR^AM$ with $E_a$= 19.04 kcal/mol precedes reactions $R^AM^\bullet_{n+1}$ with smaller $E_a$ , which does not allow the FRP of styrene to propagate when the monomer-radical is formed. Apparently, here we again meet the case when not only $E_a$ energies, but a pre-exponential factor $A$ in Eq. (1) should be taken into account. However, it is possible to confidently conclude that for styrene monomers, basing on both empirical and virtual realities,

$$k^S_{rm} > k_p. \qquad (4)$$

When suggesting that all the ERs are superpositional and occur independently on each other, input of small addition of $C_{60}$ to the discussed reaction medium means the input of all the interrelations concerning the fullerene discussed above. Since *TEMPO* cannot influence the formation of monomer-radical $FM^\bullet$ and $R^AM^\bullet$ as well as the relationship between the rate constants of the corresponding ERs, the ratio of $k^S_{FM}$ and $k^F_p$ decides the fate of the beginning of the FRCP of styrene with *TEMPO* and $C_{60}$. Virtual data do not allow evaluating the requested ratio, while experiment evidences in favor of the $SFM$ ER. Actually, attentive analysis shows that curve 2' in Figure 7b consists of three parts. The first coincides with the IP period of the FRCP of styrene with *TEMPO* (curve 1') and evidences that the rate constant $k^S_{FM}$ is the biggest. The IP is ended when all the *TEMPO* molecules are consumed that is why its length coincides with that one related to curve 1'. When reaction $SFM$ is terminated, the reaction $FM^\bullet_{n+1}$ starts just reproducing curve 2. The corresponding conversion is continued with the reaction $R^AM^\bullet_{n+1}$, so that curve 2' copies curve 2, but shifted along the axe on the length of the IP provided with *TEMPO*, analogously to the joint behavior of curves 1 and 1'. Accordingly, the rate-constant series for the FRCP of styrene with *TEMPO* and $C_{60}$, expressed in Eqs. (2)-(4), is added with one more

$$k^S_{FM} > k^F_p. \qquad (5)$$

As mentioned earlier, any of the conclusions, similar to the above made, are valid for a particular chemical contents of the polarizable medium. A set of conclusions, made in the current study and related to the styrene-based polymerization, differ considerably from previously made for methyl methacrylate [4]. Analogous consequences accompany the replacement of nitrile $AIBN$ with benzoyl-peroxide $BP$. Virtual data related to empirical results shown in Figure 7c are listed in Table 3. Analyzing these data and reasoning as above, it is easy to conclude the following. The energies $E_a$ of ERs $R^PM^\bullet$ and $R^PM^\bullet_{n+1}$, although different for $R^AM^\bullet$ and $R^AM^\bullet_{n+1}$ from Table2, are within acceptable values for the FRP of vinyl monomers. This conclusion is fully confirmed by the proximity of reference curves 1 and 1' in the figure.

The obtained $E_a$ values allow suggesting that fullerene-provided ERs form a set, rate-constant member of which follow the sequence

$$k^F_{rm} > k^F_{1m} \gg k^F_R. \qquad (6)$$

As seen in Table 3, ER (2), which determines the formation of the monomer-radical $R^PM^\bullet$, absolutely dominates, so that only reactions $R^PM^\bullet_{n+1}$ and $FR^PM$ do really compete. Since

$$k_{rm}^F > k_p, \tag{7}$$

the monomer conversion curve must include clearly seen IP, similar to that shown in Figure 7b, that is continuing until all the $C_{60}$ content is resumed, thus taking the form of a $\Gamma$ letter instead of a rounded hockey stick, that is characteristic for the FRCP of styrene with $C_{60}$, initiated with $AIBN^\bullet$. Plottings presented in Figure 7c fully confirm these expectations.

Unfortunately, experimental data for the ternary system are not available, but the data in Table 3 related to addition of *TEMPO* suggest that $SR^PM$ should be the main reaction provided with this radical. As a result, its influence on the FRCP of styrene with both $C_{60}$ and *TEMPO* will either not be noticed at all or lead to an extension in time of the IP, presented in Figure 7c, depending on how competitive is $SR^PM$ reaction with respect to $FR^PM$.

## 5. Conclusion

The concept of digital twins, which has successfully demonstrated itself in the virtualization of free radical polymerization of vinyl monomers [3,4], is used in this work to identify the reasons of the influence of minor additive of extra stable radicals on this process. The success of the implementation of the concept is provided with the reliability of the main idea, which considers polymerization as a chain reaction consisting of many independent ERs occurring in superposition. From this viewpoint, polymerization is the product of won kinetic competition in the environment of a set of elementary reactions. The ERs' final products represent the set of digital twins discussed in this article. A matrix type of tabulation is proposed to present the results obtained. The tables, supplemented with the DTs graphical equilibrium structures, represent a kind of 'passports of polymerization', the kinetics of which is under consideration. The quantum chemical approximation used assumes that the activation energy of the DTs can be considered as a reliable kinetic descriptor.

In this work, the family of vinyl monomers is presented by styrene, which undergoes rapid polymerization when initiated by the free radicals of either alkyl-nitrile $AIBN^\bullet$ or benzoyl-peroxide $BP^\bullet$. *TEMPO* and fullerene $C_{60}$ present stable radicals. Both stimulate the copolymerization with styrene, which has a significant effect on the polymerization kinetics. The FRP of the monomer and its FRCP with stable radicals are separated into a ERs' network, each of which was analyzed numerically using a standard reaction theory energy plot. Thus obtained results filled all fields of the passports mentioned above.

The first analysis of the FRCP of styrene with $C_{60}$ revealed that this process is controlled by three reactions fixate on fullerene. They are the formation of the $FM^\bullet$ monomer-radical, the capture of the $RM^\bullet$ monomer-radical $FRM$, and the capture of the free radical by the fullerene $FR$. It has been established that the rates of these reactions form a chain, the order of the rates arrangement in which depends on the chemical composition of the reaction solution. Thus, in the case of FRSP of styrene initiated by $AIBN^\bullet$, the first place is taken by the formation of the $FM^\bullet$ monomer-radical, which initiates the polymerization of styrene on fullerene. A careful analysis of the available experimental data shows that such a reaction actually occurs. However, if you replace the free radical $AIBN^\bullet$ with benzoyl-peroxide $BP^\bullet$, the capture of the $RM^\bullet$ monomer-radical by fullerene comes first in this chain. In this case, the polymerization of styrene becomes impossible, because of which an induction period with zero amplitude should be observed on the conversion curve until fullerene is completely consumed. This is exactly the type of monomer conversion under these conditions. The effect of *TEMPO* on the FRCP of styrene is reduced to a single reaction that determines the capture of the active monomer-radical, which does not depend on the chemical composition of the participants in the polymerization process.

These results were obtained by considering the "passport" data of the polymerization described above. A set of matrix data, supplemented by graphic images of the corresponding DTs, can be recommended as a mandatory document for studying the kinetics of a polymerizable medium. Available fast computational methods, which allow the consideration of radical reactions, are able to create such a passport for any reaction solution involving, for example, a wide range of vinyl and allyl monomers. The presence of such a passport can significantly reduce the development time of new polymers, cutting off unnecessary ones and leaving only a limited number of elementary reactions following from the passport.

**Acknowledgments.** This paper has been supported by the RUDN University Strategic Academic Leadership Program.